\documentclass[aps]{revtex4}

\usepackage{bm}
\usepackage{amsmath,amssymb}

\begin{document}

\title{Hamiltonian reductions of the one-dimensional Vlasov equation using phase-space moments}

\author{C. Chandre, M. Perin}
\affiliation{Centre de Physique Th\'eorique, CNRS -- Aix-Marseille Universit\'e, Campus de Luminy, case 907, 13009 Marseille, France}

\begin{abstract}
We consider Hamiltonian closures of the Vlasov equation using the phase-space moments of the distribution function. We provide some conditions on the closures imposed by the Jacobi identity. We completely solve some families of examples. As a result, we show that imposing that the resulting reduced system preserves the Hamiltonian character of the parent model shapes its phase space by creating a set of Casimir invariants as a direct consequence of the Jacobi identity. 
\end{abstract}

%\pacs{}

\maketitle

\section{Introduction}

We consider the dynamics of a beam of charged particles described by its distribution $f$ in phase space. Under the assumption that the collisions are neglected, this dynamics is described by the Vlasov equation for the distribution $f({\bf x},{\bf v};t)$: 
$$
\frac{\partial f}{\partial t}+{\bf v}\cdot \frac{\partial f}{\partial {\bf x}}+{\bf\rm F}\cdot \frac{\partial f}{\partial {\bf v}}=0,
$$
where $\bf\rm F$ is the force driving the particles, potentially of electromagnetic nature. In realistic plasma configurations, the numerical integration of such an equation is challenging. In addition, it provides information in phase space which is sometimes difficult to interpret physically. This kind of models lends itself very well to reduced models which display the physical phenomena with much more clarity (to the expense of loosing quantitative agreement in general). There has been a rather extensive emphasis on fluid reductions which means that the dynamical variables are functions of the position ${\bf x}\in {\mathbb R}^D$ only, no more on the velocity ${\bf v}\in {\mathbb R}^D$. This is for instance obtained by considering fluid moments of the distribution function
\begin{equation}
\label{eq:FM}
P_{\bf n}=\int {\rm d}^D v~  {\bf v}^{\otimes {\bf n}} f({\bf x},{\bf v};t),
\end{equation}
where ${\bf n}\in {\mathbb N}^D$ and ${\bf v}^{\otimes {\bf n}}=\prod_{i=1}^D v_i^{n_i}$. 
The resulting fluid equations are much more tractable than kinetic equations like the Vlasov equations, and are widely used in fluid and plasma physics. The difficulty in using these fluid equations in plasma physics is that they do not sufficiently take into account kinetic effects. In order to remedy this problem, there is a need to include a high number of fluid moments which, as a result, complexifies the analysis of the resulting physical phenomena, since it involves a high number of coupled nonlinear partial differential equations. 

An alternative approach is to consider phase space moments for which the reduced models become finite dimensional (i.e., described by ordinary differential equations) which facilitates the analysis. The phase space moments are defined as
\begin{equation}
\label{eq:PSM}
P_{{\bf n},{\bf m}}=\iint {\rm d}^D x{\rm d}^D v~ {\bf x}^{\otimes {\bf n}} {\bf v}^{\otimes {\bf m}} f({\bf x},{\bf v};t),
\end{equation}
where $({\bf n},{\bf m})\in {\mathbb N}^{2D}$. 
We notice that $P_{0,0}$ is the total number of particles and $P_{{\bf n},0}/P_{0,0}$, with $n_i=\delta_{i,j}$, is the $j$th component of the average distance (also called beam centroid) and $P_{0,{\bf m}}/P_{0,0}$ with $m_i=\delta_{i,j}$, is the $j$th component of the average velocity associated with the distribution function. Some of these phase space moments are directly measurable quantities which make them appropriate variables to describe the dynamics.
There is a rather extensive literature of phase space moment reductions, especially in the field of beam dynamics in linear plasma-based accelerators~\cite{chan83,berz85,lyse88,chan90,holm90,drag92,scov94,chan95,lyse95,lyse97,shad99,shad10,shad12,lee12,evst13}. 

The principle of the reduction is the following one: A reduced set of phase space moments $P_{{\bf n},{\bf m}}$, relevant to the problem at hand, has to be identified. We denote the set of indices $({\bf n},{\bf m})$ of these relevant moments ${\cal V}\subset {\mathbb N}^{2D}$. In general, the dynamics of these relevant moments depend on additional moments whose set of indices is denoted ${\cal C}\subset {\mathbb N}^{2D}$. The closure on $P_{{\bf n},{\bf m}}$ for $({\bf n},{\bf m})\in {\cal C}$ is a specific function of the dynamical variables $P_{{\bf k},{\bf l}}$ for $({\bf k},{\bf l})\in {\cal V}$ which preserves the Hamiltonian properties of the parent model, i.e., the Vlasov equation. The dynamics of any observable $F$, scalar function of the variables $P_{{\bf n},{\bf m}}$ for $({\bf n},{\bf m})\in {\cal V}$, is described by a Hamiltonian ${\cal H}$ and a Poisson bracket $\{\cdot,\cdot\}$ as
$$
\frac{dF}{dt}=\{F,{\cal H}\},
$$
and the Poisson bracket is a bilinear operator satisfying the antisymmetry property
$$
\{F,G\}=-\{G,F\},
$$
the Leibniz rule,
$$
\{FG,H\}=F\{G,H\}+\{F,H\}G,
$$
and the Jacobi identity,
$$
\{F,\{G,H\}\}+\{H,\{F,G\}\}+\{G,\{H,F\}\}=0,
$$
for any observables $F$, $G$ and $H$. 
As usual, the last property is the crucial one and also the most difficult one to satisfy. It leads to a very specific subset of possible closures. As a result, it drastically constrains the phase space of the reduced dynamics. In particular, the Jacobi identity imposes the foliation of phase space by a set of Casimir invariants which are observables $C$ which Poisson-commute with all the other observables, i.e., $\{C,F\}=0$ for all observables $F$. For an introduction on Hamiltonian systems and Poisson algebras, we refer to Refs.~\cite{mars02,morr98}.

Some Hamiltonian fluid closures were investigated in Refs.~\cite{peri15a,peri15b} using the fluid moments defined by Eq.~(\ref{eq:FM}). In particular it was noticed that some $N$-moment fluid closures were obtained using the waterbag distribution functions or by dimensional analysis by solving the Jacobi identity. One of the objectives of the present work is to follow the same procedure for the phase space moments defined by Eq.~(\ref{eq:PSM})  

In Sec.~\ref{sec:def}, we recall the Poisson bracket for the one-dimensional Vlasov equation and identify some basic Poisson sub-algebras. We also introduce central moments which are very practical to work with. In Sec.~\ref{sec:close}, we provide some general results on the possible closures, and list some examples, by considering specific subsets of variables.   

\section{Definitions and model}
\label{sec:def}

\subsection{Poisson subalgebras}
\label{sec:sub}

We consider the one-dimensional ($D=1$) Vlasov equation 
$$
\frac{\partial f}{\partial t}=-v\frac{\partial f}{\partial x}-{\rm F} \frac{\partial f}{\partial v}. 
$$
The dynamical field variable is $f(x,v)$ and we consider the algebra of functionals of $f$.  
The Poisson bracket is given by
\begin{equation}
\label{eq:PBV}
\{F,G\}=\iint {\rm d}x {\rm d}v f\left(\frac{\partial}{\partial x}\frac{\delta F}{\delta f}\frac{\partial }{\partial v}\frac{\delta G}{\delta f}-\frac{\partial}{\partial x}\frac{\delta G}{\delta f}\frac{\partial}{\partial v}\frac{\delta F}{\delta f}
\right),
\end{equation}
where $\delta F/\delta f$ is the functional derivative of $F$ with respect to $f$. We recall that the Casimir invariants of the bracket~(\ref{eq:PBV}) are given by 
$$
C=\iint {\rm d}x{\rm d} v ~ \gamma(f),
$$
for any scalar function $\gamma$.

The choice of the Hamiltonian depends on the problem at hand. For instance, for the one-dimensional Vlasov-Poisson equation for which ${\rm F}=\partial \varphi/\partial x$ and $\Delta \varphi=\int {\rm d}v f$, the Hamiltonian is given by
\begin{equation}
\label{eq:H}
H[f]=\frac{1}{2}\iint {\rm d}x{\rm d} v~ \left(v^2 f +f{\cal L} f\right),
\end{equation}
where ${\cal L}f=\Delta^{-1} \int f {\rm d}v $ (defined using proper boundary conditions). However the specific expression of the Hamiltonian is not particularly relevant for the analysis presented below. In particular the expression of the Hamiltonian as a function of the phase space moments does not present any major difficulty. It might involve a truncation in order to have the Hamiltonian in the finite-dimensional Poisson algebra, but this truncation doe not affect the Poisson structure of the resulting truncated model. On the contrary, truncating the Poisson bracket results in general in the break-up of the Hamiltonian structure of the model, since the Jacobi identity is in general not satisfied. As we shall see below, this drastically affects the  resulting dynamics.
 
Expressed in the phase space moments, i.e., moving from functionals of $f$ to functions of $P_{n,m}$ defined as
$$
P_{n,m}=\iint {\rm d}x {\rm d}v~ x^n v^m f(x,v), 
$$ 
the Poisson bracket becomes~\cite{holm90,chan90,gibb08,gayb12}
\begin{equation}
\{F,G\}=(nl-mk)P_{n+k-1,m+l-1}\frac{\partial F}{\partial P_{n,m}}\frac{\partial G}{\partial P_{k,l}}, \label{eq:PB_P}
\end{equation}
with implicit summation over repeated indices $(n,m,k,l)$. 

We notice that $P_{0,0}$ (which is the total number of particles) is a Casimir invariant of the Poisson bracket~(\ref{eq:PB_P}). In what follows, we assume $P_{0,0}=1$ without loss of generality, by rescaling the distribution function by the total number of particles.
First we identify several Poisson sub-algebras:
\begin{itemize}
\item Subalgebra 1: the set of observables $F(P_{1,0},P_{0,1})$. The Poisson bracket is the canonical one 
$$
\{F,G\}_1=\frac{\partial F}{\partial P_{1,0}}\frac{\partial G}{\partial P_{0,1}}-\frac{\partial F}{\partial P_{0,1}}\frac{\partial G}{\partial P_{1,0}}.
$$
The coordinate centroid is canonically conjugate to the momentum centroid which is very reasonable from a macroscopic viewpoint. 
\item Subalgebra 2: the set of observables $F(P_{1,0},P_{0,1},P_{1,1})$. The Poisson bracket is
$$
\{F,G\}_{1,1}=\{F,G\}_1+P_{1,0}\left(\frac{\partial F}{\partial P_{1,0}}\frac{\partial G}{\partial P_{1,1}}-\frac{\partial F}{\partial P_{1,1}}\frac{\partial G}{\partial P_{1,0}} \right)+P_{0,1}\left(\frac{\partial F}{\partial P_{1,1}}\frac{\partial G}{\partial P_{0,1}}-\frac{\partial F}{\partial P_{0,1}}\frac{\partial G}{\partial P_{1,1}} \right).
$$
This Poisson bracket has one Casimir invariant $C=P_{1,1}-P_{1,0}P_{0,1}$, the covariance of $f$ with respect to $x$ and $v$. This subalgebra is not particularly interesting since it reduces the dynamics to one degree of freedom [the macroscopic degree of freedom $(P_{1,0},P_{0,1})$] on each leaf defined by the value of the Casimir invariant. Like Subalgebra 1, it does not possess any degree of freedom related to the shape of the beam. 
\item Subalgebra 3: the set of observables $F(P_{1,0},P_{0,1},P_{1,1},P_{2,0},\ldots,P_{N,0})$ for any arbitrary $N\geq 2$. The Poisson bracket is
$$
\{F,G\}_{N,0}=\{F,G\}_{1,1}+nP_{n-1,0}\left(\frac{\partial F}{\partial P_{n,0}}\frac{\partial G}{\partial P_{0,1}}-\frac{\partial F}{\partial P_{0,1}}\frac{\partial G}{\partial P_{n,0}} \right)
+nP_{n,0}\left(\frac{\partial F}{\partial P_{n,0}}\frac{\partial G}{\partial P_{1,1}}-\frac{\partial F}{\partial P_{1,1}}\frac{\partial G}{\partial P_{n,0}} \right),
$$
where the implicit summation over $n$ goes from 2 to $N$. This Poisson bracket has $N-2$ Casimir invariants $C_n$ given by
$$
C_n=\frac{1}{(P_{2,0}-P_{1,0}^2)^{n/2}}\sum_{k=0}^n (-1)^{k}\binom{n}{k} P_{1,0}^k P_{n-k,0},
$$
for $n=3,\ldots, N$. 
We notice that the first three subalgebras do not contain the kinetic energy $\int {\rm d}x {\rm d}v f v^2/2=P_{0,2}/2$. Indeed Hamiltonian~(\ref{eq:H}) is rewritten as
\begin{equation}
\label{eq:H2}
H=\frac{P_{0,2}}{2}+V(\{P_{n,0}\}_{n\geq 0}),
\end{equation}
where $V$ is an arbitrary function. 
Therefore, in order to close the system, we need to express $P_{0,2}$ as a function of the other variables. A common way of doing it is to decompose $P_{0,2}/2$ by $P_{0,1}^2/2$ (the macroscopic kinetic energy) and a fluctuating term corresponding to an internal energy $U$ which is a well-chosen function of all the other dynamical variables (see the example provided in Subalgebra 4), accounting for the microscopic dynamics. Subalgebra 3 is particularly interesting since it allows the treatment of arbitrary nonlinearities in the potential created, e.g., by the elements of a particle accelerator, in a closed form. In addition, given the number of dynamical variables and the number of Casimir invariants, the system is equivalent to a two degrees of freedom Hamiltonian system in terms of effective phase space dimension: There is one macroscopic degree of freedom associated with the position of the beam, namely $(P_{1,0},P_{0,1})$, and one microscopic degree of freedom associated with the shape of the beam. 

The corresponding bracket $\{\cdot,\cdot\}_{2,0}$ in the variables $(P_{1,0},P_{0,1},P_{2,0},P_{1,1})$ is non-canonical. In order to make the bracket canonical, we perform the non-canonical change of variables 
$$
(P_{1,0},P_{0,1},P_{2,0},P_{1,1})\mapsto \left( P_{1,0},P_{0,1},\sqrt{P_{2,0}-P_{1,0}^2},\frac{P_{1,1}-P_{1,0}P_{0,1}}{\sqrt{P_{2,0}-P_{1,0}^2}}\right).
$$   
We notice that for non-zero $f$, the variance $P_{2,0}-P_{1,0}^2$ is always strictly positive. The last two variables in the above expression are the two canonically conjugate variables describing the degree of freedom associated with the shape of the beam: $\sqrt{P_{2,0}-P_{1,0}^2}$ is the standard deviation of the beam and $(P_{1,1}-P_{1,0}P_{0,1})/\sqrt{P_{2,0}-P_{1,0}^2}$ its canonically conjugate momentum.  

There is also another subalgebra which is related to this one by inverting the roles of $x$ and $v$. The Poisson bracket is the same as the one above up to a change of sign. 
 
\item Subalgebra 4: the set of observables $F(P_{1,0},P_{0,1},P_{1,1},P_{2,0},P_{0,2})$. The Poisson bracket is
\begin{eqnarray*}
\{F,G\}_2&=&\{F,G\}_{1,1}+2P_{1,0}\left(\frac{\partial F}{\partial P_{2,0}}\frac{\partial G}{\partial P_{0,1}}-\frac{\partial F}{\partial P_{0,1}}\frac{\partial G}{\partial P_{2,0}} \right)
+2P_{0,1}\left(\frac{\partial F}{\partial P_{1,0}}\frac{\partial G}{\partial P_{0,2}}-\frac{\partial F}{\partial P_{0,2}}\frac{\partial G}{\partial P_{1,0}} \right) \nonumber\\
&& +2P_{2,0}\left(\frac{\partial F}{\partial P_{2,0}}\frac{\partial G}{\partial P_{1,1}}-\frac{\partial F}{\partial P_{1,1}}\frac{\partial G}{\partial P_{2,0}}\right)+2P_{0,2}\left(\frac{\partial F}{\partial P_{1,1}}\frac{\partial G}{\partial P_{0,2}}-\frac{\partial F}{\partial P_{0,2}}\frac{\partial G}{\partial P_{1,1}} \right)\nonumber \\
&& +4P_{1,1}\left(\frac{\partial F}{\partial P_{2,0}}\frac{\partial G}{\partial P_{0,2}}-\frac{\partial F}{\partial P_{0,2}}\frac{\partial G}{\partial P_{2,0}} \right).
\end{eqnarray*}
This subalgebra and this bracket correspond to the ones identified in Refs.~\cite{shad99,shad10,shad12}. The Poisson bracket has one Casimir invariant given by
$$
C=(P_{2,0}-P_{1,0}^2)(P_{0,2}-P_{0,1}^2)-(P_{1,1}-P_{1,0}P_{0,1})^2,
$$
which is called the emittance in accelerator physics. It corresponds to the determinant of the covariance matrix. 
This subalgebra allows the inclusion of the full kinetic energy $P_{0,2}/2$, even though it is restricted to quadratic potential (i.e., linear forces). This subalgebra is not fundamentally different from Subalgebra 3 since both of them have only one degree of freedom associated with the shape of the beam. Given a value of the Casimir invariant, both subalgebras lead to the same dynamics provided the kinetic term $P_{0,2}/2$ is replaced by 
\begin{equation}
\label{eq:K2}
\frac{P_{0,2}}{2}=\frac{P_{0,1}^2}{2}+\frac{(P_{1,1}-P_{1,0}P_{0,1})^2+C}{2(P_{2,0}-P_{1,0}^2)},
\end{equation}
where the second term on the right hand side of the above equation is the internal energy. In particular, the consequence is that the resulting Hamiltonian, derived from Hamiltonian~(\ref{eq:H2}) using Eq.~(\ref{eq:K2}), presents an explicit (quadratic) dependence on $P_{1,1}-P_{1,0}P_{0,1}$ which is the canonically conjugate momentum of the standard deviation of the distribution function. 
\item Subalgebra 5: the set of observables $F(\{P_{n,m}\}_{m\leq 1})$. This is not a finite-dimensional subalgebra since it contains an infinite set of variables. This subalgebra corresponds to the subalgebra in the fluid case where the observables are functions of the fluid density and the fluid velocity only. The Poisson bracket writes
$$
\{F,G\}=nP_{n+k-1,0}\left(\frac{\partial F}{\partial P_{n,0}}\frac{\partial G}{\partial P_{k,1}}-\frac{\partial F}{\partial P_{k,1}}\frac{\partial G}{\partial P_{n,0}}\right)+(n-k)P_{n+k-1,1}\frac{\partial F}{\partial P_{n,1}}\frac{\partial G}{\partial P_{k,1}},
$$
with the implicit summation over $n$ and $k$ from 0 to infinity. 
This Poisson bracket has $P_{0,0}$ as Casimir invariant. In particular we notice that Subalgebra 3 is a Poisson subalgebra of Subalgebra 5.  The main drawback of this subalgebra is that it does not have the second order moment in velocity, $P_{0,2}$, linked to the kinetic energy term in the Hamiltonian, among its variables. 
\end{itemize} 
The Casimir invariants associated with the above families of subalgebras, in particular Subalgebra 3, evidence a particularly well suited change of variables, the central moments which will be considered in what follows. These Poisson subalgebras are the most obvious ones, but not the only ones. In the next section, we identify additional non-trivial Poisson subalgebras which are more obvious in the other set of variables, the central moments. 

\subsection{Phase space central moments}

We introduce the phase-space central moments defined by
$$
W_{n,m}=\iint {\rm d}x {\rm d}v~ (x-P_{1,0})^n(v-P_{0,1})^m f(x,v). 
$$
We notice that the covariant matrix elements are $W_{2,0}$, $W_{0,2}$ and $W_{1,1}$. 
The expressions of $W_{n,m}$ with respect to $P_{n,m}$ are given by
$$
W_{n,m}=\sum_{k=0}^n\sum_{l=0}^m (-1)^{n+m-k-l}\binom{n}{k}\binom{m}{l}P_{1,0}^{n-k}P_{0,1}^{m-l}P_{k,l}. 
$$
This expression is invertible~:
$$
P_{n,m}=\sum_{k=0}^n\sum_{l=0}^m \binom{n}{k}\binom{m}{l}P_{1,0}^{n-k}P_{0,1}^{m-l}W_{k,l},
$$
where $W_{0,0}=1$ and $W_{1,0}=W_{0,1}=0$. 
The expression of the Poisson bracket~(\ref{eq:PB_P}) is 
\begin{equation}
\label{eq:brackW}
\{F,G\}=\frac{\partial F}{\partial P_{1,0}}\frac{\partial G}{\partial P_{0,1}}-\frac{\partial F}{\partial P_{0,1}}\frac{\partial G}{\partial P_{1,0}}+\alpha_{nmkl}\frac{\partial F}{\partial W_{n,m}}\frac{\partial G}{\partial W_{k,l}},
\end{equation}
with 
\begin{equation}
\label{eq:alpha}
\alpha_{nmkl}=(nl-mk)W_{n+k-1,m+l-1}+km W_{n,m-1}W_{k-1,l}-nl W_{n-1,m}W_{k,l-1}.
\end{equation}
The implicit summation in Eq.~(\ref{eq:brackW}) excludes $(n,m)$ (or $(k,l)$) equal to $(0,0)$, $(1,0)$ and $(0,1)$.  In order to obtain the expression of Bracket~(\ref{eq:brackW}), we used the Vandermonde's identity. 
We notice that the tensor $\alpha$ is symmetric, in the sense that $\alpha_{nmkl}=\alpha_{klnm}$ in order to ensure the antisymmetry of the Poisson bracket.  

The expression of the Poisson bracket in this set of variables is particularly interesting since it decouples the macroscopic degree of freedom $(P_{1,0},P_{0,1})$ from the microscopic degrees of freedom, represented by the variables $W_{n,m}$, regulating the shape of the beam. 
In particular, the set of observables $F(\{W_{n,m}\})$ is a Poisson subalgebra, since the coefficient of $\alpha$ do not depend on $P_{1,0}$ and $P_{0,1}$. 

We notice that the coefficients in Eq.~(\ref{eq:alpha}) contain a linear part in the $W$s whose expression is identical to the one of the bracket in $P$s in Eq.~(\ref{eq:PB_P}). However a noticeable difference between the bracket expressed in the variables $P$s with the one expressed in the variables $W$s is that the bracket in $W$s contains quadratic terms which drastically complicate the analysis and the reduction procedure as we shall see below.

\section{Hamiltonian closures}
\label{sec:close}
The problem of the closure is to select a {\em finite} number of moments $P_{n,m}$ for the problem at hand, i.e., select a finite subset ${\cal V}$ of ${\mathbb N}^2$ in which $P_{n,m}$ is a variable when $(n,m)\in {\cal V}$. When the set of observables $F$ depending on these relevant variables are considered, the Poisson bracket explicitly depends on a finite number of additional $P_{n,m}$ which are not considered as relevant dynamical variables. The set of these additional phase space moments is denoted ${\cal C}$. Therefore the bracket has to be closed, i.e., for all $(n,m)\in {\cal C}$, we assume that $$P_{n,m}=P_{n,m}(\{P_{k,l}\}_{(k,l)\in {\cal V}}).$$

\subsection{Finite-dimensional Lie-Poisson algebras}
\label{sec:FDLP}

Finite dimensional Lie-Poisson approximations of the infinite dimensional Lie-Poisson structure of the Vlasov equation have been proposed in the literature~\cite{drag88,chan90,drag92,scov94,chan95}. It was shown that the set of moments $P_{n,m}$ with $M\leq n+m\leq N$ is a Poisson algebra if $P_{n,m}=0$ for all $n+m>N$ and $M\geq 2$. In particular, this reduction removes ad hoc the macroscopic degree of freedom $(P_{1,0},P_{0,1})$, hypothesizing that this macroscopic degree of freedom is decoupled from the internal degrees of freedom regulating the shape of the beam. 

These closures leading to finite dimensional Lie-Poisson algebras were used to devise numerical algorithms to integrate the Vlasov equation using phase-space moments of the Vlasov distribution~\cite{chan95}, using the symplectic integration algorithms developed for finite dimensional Lie-Poisson algebras. The main ingredient behind these finite-dimensional Lie-Poisson algebras is the linearity of the coefficients in Bracket~(\ref{eq:PB_P}). 
 
The simplest example is when the dynamical variables are $(P_{0,2},P_{1,1},P_{2,0})$ (i.e., $N=M=2$). The Poisson bracket writes 
$$
\{F,G\}_{LP1}=2P_{2,0}\left(\frac{\partial F}{\partial P_{2,0}}\frac{\partial G}{\partial P_{1,1}}-\frac{\partial F}{\partial P_{1,1}}\frac{\partial G}{\partial P_{2,0}} \right)
+2P_{0,2}\left(\frac{\partial F}{\partial P_{1,1}}\frac{\partial G}{\partial P_{0,2}}-\frac{\partial F}{\partial P_{0,2}}\frac{\partial G}{\partial P_{1,1}} \right)+4P_{1,1}\left(\frac{\partial F}{\partial P_{2,0}}\frac{\partial G}{\partial P_{0,2}}-\frac{\partial F}{\partial P_{0,2}}\frac{\partial G}{\partial P_{2,0}} \right).
$$
In this case, there is no need to close the system since the set of observables of $(P_{0,2},P_{1,1},P_{2,0})$ is a Poisson subalgebra (which is actually a subalgebra of Subalgebra 4 in Sec.~\ref{sec:sub}). It has the beam emittance $P_{2,0}P_{0,2}-P_{1,1}^2$ as Casimir invariant. The first non-trivial case corresponds to a system whose dynamical variables are $(P_{0,2},P_{1,1},P_{2,0},P_{0,3},P_{1,2},P_{2,1},P_{3,0})$, i.e., $M=2$ and $N=3$. The Poisson bracket is given by
\begin{eqnarray}
\{F,G\}&=&\{F,G\}_{LP1}+3P_{3,0}\left(\frac{\partial F}{\partial P_{3,0}}\frac{\partial G}{\partial P_{1,1}}-\frac{\partial F}{\partial P_{1,1}}\frac{\partial G}{\partial P_{3,0}} \right)-3P_{0,3}\left(\frac{\partial F}{\partial P_{0,3}}\frac{\partial G}{\partial P_{1,1}}-\frac{\partial F}{\partial P_{1,1}}\frac{\partial G}{\partial P_{0,3}} \right)\nonumber \\
&& +2P_{3,0}\left(\frac{\partial F}{\partial P_{2,0}}\frac{\partial G}{\partial P_{2,1}}-\frac{\partial F}{\partial P_{2,1}}\frac{\partial G}{\partial P_{2,0}} \right)-2P_{0,3}\left(\frac{\partial F}{\partial P_{0,2}}\frac{\partial G}{\partial P_{1,2}}-\frac{\partial F}{\partial P_{1,2}}\frac{\partial G}{\partial P_{0,2}} \right)\nonumber\\
&& +6P_{2,1}\left(\frac{\partial F}{\partial P_{3,0}}\frac{\partial G}{\partial P_{0,2}}-\frac{\partial F}{\partial P_{0,2}}\frac{\partial G}{\partial P_{3,0}} \right)-6P_{1,2}\left(\frac{\partial F}{\partial P_{0,3}}\frac{\partial G}{\partial P_{2,0}}-\frac{\partial F}{\partial P_{2,0}}\frac{\partial G}{\partial P_{0,3}} \right)\nonumber \\
&&+4P_{2,1}\left(\frac{\partial F}{\partial P_{2,0}}\frac{\partial G}{\partial P_{1,2}}-\frac{\partial F}{\partial P_{1,2}}\frac{\partial G}{\partial P_{2,0}} \right)-4P_{1,2}\left(\frac{\partial F}{\partial P_{0,2}}\frac{\partial G}{\partial P_{2,1}}-\frac{\partial F}{\partial P_{2,1}}\frac{\partial G}{\partial P_{0,2}} \right)\nonumber\\
&&+P_{1,2}\left(\frac{\partial F}{\partial P_{1,1}}\frac{\partial G}{\partial P_{1,2}}-\frac{\partial F}{\partial P_{1,2}}\frac{\partial G}{\partial P_{1,1}} \right)-P_{2,1}\left(\frac{\partial F}{\partial P_{1,1}}\frac{\partial G}{\partial P_{2,1}}-\frac{\partial F}{\partial P_{2,1}}\frac{\partial G}{\partial P_{1,1}} \right)+\{F,G\}_4, \label{eq:LP4}
\end{eqnarray}
where 
\begin{eqnarray*}
\{F,G\}_4&=& -3P_{0,4}\left(\frac{\partial F}{\partial P_{0,3}}\frac{\partial G}{\partial P_{1,2}}-\frac{\partial F}{\partial P_{1,2}}\frac{\partial G}{\partial P_{0,3}} \right)-3P_{4,0}\left(\frac{\partial F}{\partial P_{2,1}}\frac{\partial G}{\partial P_{3,0}}-\frac{\partial F}{\partial P_{3,0}}\frac{\partial G}{\partial P_{2,1}} \right)\\
&& -6P_{1,3}\left(\frac{\partial F}{\partial P_{0,3}}\frac{\partial G}{\partial P_{2,1}}-\frac{\partial F}{\partial P_{2,1}}\frac{\partial G}{\partial P_{0,3}} \right)-6P_{3,1}\left(\frac{\partial F}{\partial P_{1,2}}\frac{\partial G}{\partial P_{3,0}}-\frac{\partial F}{\partial P_{3,0}}\frac{\partial G}{\partial P_{1,2}} \right)\\
&& -9P_{2,2}\left(\frac{\partial F}{\partial P_{0,3}}\frac{\partial G}{\partial P_{3,0}}-\frac{\partial F}{\partial P_{3,0}}\frac{\partial G}{\partial P_{0,3}} \right)-3P_{2,2}\left(\frac{\partial F}{\partial P_{1,2}}\frac{\partial G}{\partial P_{2,1}}-\frac{\partial F}{\partial P_{2,1}}\frac{\partial G}{\partial P_{1,2}} \right). 
\end{eqnarray*}
The finite-dimensional Lie-Poisson argument consists in setting $P_{4,0}=P_{3,1}=P_{2,2}=P_{1,3}=P_{0,4}=0$, i.e., removing $\{\cdot,\cdot\}_4$ from Bracket~(\ref{eq:LP4}). The resulting bracket~(\ref{eq:LP4}) satisfies the Jacobi identity. It has one Casimir invariant:
$$
C=P_{0,3}^2P_{3,0}^2-3P_{1,2}^2P_{2,1}^2+4P_{2,1}^3P_{0,3}+4P_{1,2}^3P_{3,0}-6P_{1,2}P_{2,1}P_{0,3}P_{3,0},$$
which is the equivalent of the beam emittance for the case $N=M=2$.  

More generally, it can be shown that the Jacobi identity restricted to the observables $F(\{P_{n,m}\}_{M\leq n+m\leq N})$ is a linear function of $P_{n,m}$ with $n+m>N$, for all $N\geq M\geq 2$. By setting $P_{n,m}=0$ for $n+m>N$, the resulting bracket satisfies the Jacobi identity. In fact, only those closures result in a Poisson bracket, so $P_{n,m}=0$ for $n+m>N$ is a necessary and sufficient condition for the truncated system to be Hamiltonian.

However, the truncation $P_{n,m}=0$ for $n+m>N$ might not be a reasonable approximation of the Vlasov dynamics even though it is a truncation which leads to a Hamiltonian system. As a matter of fact, the construction of the truncated system relies on an ad hoc assumption of decoupling between the macroscopic and the microscopic degrees of freedom which is in general not valid. In the variables $P$s, these degrees of freedom are naturally coupled. We have seen above that a more rigorous way of decoupling these degrees of freedom is to move to the central moments. 
Restricting the relevant dynamical variables to $P_{n,m}$ with $M\leq n+m\leq N$, even though practical from a theoretical perspective, might not be optimal to model/approximate the dynamics of the Vlasov equation with a finite set of variables. In principle, it would be best to first decouple the macroscopic degree of freedom from the microscopic ones, i.e., moving to the phase space central moments, before applying the reduction procedure. However the main obstacle is that the coefficients in the bracket are no longer linear in the dynamical variables, so the above Lie-Poisson reduction cannot be applied. 

The strategy we adopt in this article is different: First, the set of relevant variables is chosen, e.g., by taking into account the physical problem at hand. In other terms, a finite subset ${\cal V}$ of ${\mathbb N}^2$ is given so that $P_{n,m}$ belongs to the relevant dynamical variables for $(n,m)\in {\cal V}$. The expression of Bracket~(\ref{eq:PB_P}) shows that it explicitly depends on a finite number of moments (whose indices belong to a finite subset ${\cal C}$ of ${\mathbb N}^2$) which do not belong to the dynamical variables. These functions need to be closed, i.e., one needs to determine the functions $P_{n,m}(\{P_{k,l}\}_{(k,l)\in {\cal V}})$ for $(n,m)\in {\cal C}$, in order to obtain a finite-dimensional algebra. It should be noticed that, in general, these closures do not satisfy the Jacobi identity, and therefore do not lead to Poisson algebras. Here we determine the specific conditions on the closures in order to satisfy the Jacobi identity and lead to a reduced system which is of the Poisson type. In particular, taking higher order $P$s equal to zero does not lead in general to a Poisson algebra. Some examples are shown in the following sections.      

In order to simplify the calculations, the general idea is to decouple the location from the shape of a lump of density in phase space in the spirit of Ref.~\cite{scov94}. To this purpose, we use the central moments as dynamical variables. The problem of closure is now to find the closures 
$$W_{n,m}=W_{n,m}(P_{1,0},P_{0,1},\{W_{k,l}\}_{(k,l)\in {\cal V}}),$$
for all $(n,m)\in {\cal C}$, leading to a Poisson algebra. 
From a physical point of view, it would be desirable to have at least $(0,2)\in {\cal V}$ in order to include the kinetic energy term of the Hamiltonian as a dynamical variable, and $(n,0)\in {\cal V}$ for sufficiently large $n$ in order to include higher order nonlinearities in the potential.

Given the expression of the coefficients $\alpha$ in Eq.~(\ref{eq:alpha}), we notice that the truncated bracket is always linear in the closure functions $W_{n,m}$ where $(n,m)\in {\cal C}$. 

\subsection{Jacobi identity}

We consider the following bracket:
\begin{equation}
\label{eq:brackWT}
\{F,G\}=\frac{\partial F}{\partial P_{1,0}}\frac{\partial G}{\partial P_{0,1}}-\frac{\partial F}{\partial P_{0,1}}\frac{\partial G}{\partial P_{1,0}}+\sum_{(n,m)\in {\cal V} \atop (k,l)\in {\cal V}}\alpha_{nmkl}\frac{\partial F}{\partial W_{n,m}}\frac{\partial G}{\partial W_{k,l}},
\end{equation}
which is obtained by truncating Bracket~(\ref{eq:brackW}), reducing the set of dynamical variables to a finite set of $W$s. We notice that $\alpha_{nmkl}$ depends on $W$s which are not dynamical variables. This dependence on extra $W_{n,m}$ for $(n,m)\in{\cal C}$ comes from the linear term in $\alpha$, namely $(nl-mk)W_{n+k-1,m+l-1}$. We close the system by assuming that these extra $W$s depend on the dynamical variables $P_{1,0}$, $P_{0,1}$ and $W_{k,l}$ for $(k,l)\in{\cal V}$. 

In order to satisfy the Jacobi identity, a necessary (but not sufficient) condition is to have $\alpha_{n,m,k,l}$ independent of $P_{1,0}$ and $P_{0,1}$ for all $(n,m)$ and $(k,l)$ belonging to ${\cal V}$, i.e., all the closures only depend on $W_{n,m}$ for $(n,m)\in {\cal V}$. In order to show this, we consider $F=F(P_{0,1},P_{1,0})$ and $G$ and $H$ only functions of $W$s. From $\{F,G\}=\{H,F\}=0$, the Jacobi identity reduces to imposing $\{\{G,H\},F\}=0$. Taking respectively $F=P_{1,0}$ and $F=P_{0,1}$ leads to $\alpha_{nmkl}$ independent of $P_{0,1}$ and $P_{1,0}$ respectively. This means that the decoupling between the macroscopic degree of freedom and the microscopic ones is preserved in Bracket~(\ref{eq:brackWT}) by the truncation, and directly results from the Jacobi identity. These degrees of freedom are, of course, coupled through the Hamiltonian. 

As a consequence, the Jacobi identity reduces to the following identities
\begin{equation}
\label{eq:JacA}
\alpha_{nmkl}\frac{\partial \alpha_{n'm'k'l'}}{\partial W_{nm}}+\alpha_{nmk'l'}\frac{\partial \alpha_{kln'm'}}{\partial W_{nm}}+\alpha_{nmn'm'}\frac{\partial \alpha_{k'l'kl}}{\partial W_{nm}}=0, 
\end{equation}
where the summation over $(n,m)\in{\cal V}$ is implicit, and this identity should be true for all $(n',m')$, $(k',l')$ and $(k,l)$ in ${\cal V}$. 

First we assume that $(1,1)\in{\cal V}$, i.e., $W_{1,1}$ is a dynamical variable, and we consider the case where $k=l=1$, $n'+k'=N+1$ and $m'+l'=M+1$ where $(N,M)\in {\cal C}$. Equation~(\ref{eq:JacA}) leads to 
\begin{equation}
\label{eq:rescaleW}
(n'l'-m'k')\left( (M-N)W_{N,M}+(n-m)W_{n,m}\frac{\partial W_{N,M}}{\partial W_{n,m}}\right)=0,
\end{equation}
with implicit summation over $(n,m)\in {\cal V}$ and for all $(N,M)\in {\cal C}$. 
If $n'l'-m'k'=0$, the terms proportional to $W_{N,M}$ do not appear in Bracket~(\ref{eq:brackWT}) given the expression of $\alpha$ [see Eq.~(\ref{eq:alpha})]. Therefore for $(N,M)\in{\cal C}$, we have necessarily $n'l'-m'k'\not= 0$. The solutions of Eq.~(\ref{eq:rescaleW}) are obtained by rescaling. For instance, if $(0,2)\in {\cal V}$, i.e., if $W_{0,2}$ is a dynamical variable, then the closures $W_{N,M}$ are functions of the rescaled variables $W_{n,m}W_{0,2}^{(n-m)/2}$, i.e., written as
$$
W_{N,M}=W_{0,2}^{(M-N)/2}\widetilde{W}_{N,M}\left( W_{n,m}W_{0,2}^{(n-m)/2}\right),
$$
for all $(N,M)\in {\cal C}$. If $(2,0)\in {\cal V}$, an alternative way to describe the closure is the following one:
$$
W_{N,M}=W_{2,0}^{(N-M)/2}\widetilde{W}_{N,M}\left( W_{n,m}/W_{2,0}^{(n-m)/2}\right).
$$
These are necessary conditions, but not sufficient ones. Unfortunately we have not found general expressions for the closures in a case of an arbitrary set of variables. Below we examine different sets of variables where the subalgebra of observables $F(P_{1,0},P_{0,1},\{P_{k,l}\}_{(k,l)\in {\cal V}})$ is the same as the subalgebra of observables $F(P_{1,0},P_{0,1},\{W_{k,l}\}_{(k,l)\in {\cal V}})$ for consistency. These examples illustrate the complexity of finding Hamiltonian closures, i.e., solving the Jacobi identity. On each example, we will see how the Jacobi identity suggests relevant variables and creates a certain number of Casimir invariants which foliate the phase space of the truncated Hamiltonian system. 

\subsection{Hamiltonian closures for $(P_{1,0},P_{0,1},W_{0,2},W_{2,0})$}
\label{ex:1}

The simplest example is when the dynamical variables are $(P_{1,0},P_{0,1},W_{0,2},W_{2,0})$ and the closure has to be on $W_{1,1}$. Bracket~(\ref{eq:brackWT}) reads
$$
\{F,G\}=\frac{\partial F}{\partial P_{1,0}}\frac{\partial G}{\partial P_{0,1}}-\frac{\partial F}{\partial P_{0,1}}\frac{\partial G}{\partial P_{1,0}}+4W_{1,1}(W_{2,0},W_{0,2})\left( \frac{\partial F}{\partial W_{2,0}}\frac{\partial G}{\partial W_{0,2}}-\frac{\partial F}{\partial W_{0,2}}\frac{\partial G}{\partial W_{2,0}}\right). 
$$
It is rather straightforward to check that this bracket is a Poisson bracket for any function $W_{1,1}$. A trivial remark is that if $W_{1,1}=0$, $W_{2,0}$ and $W_{0,2}$ are Casimir invariants. Therefore $W_{1,1}$ is an arbitrary function of the observables which are Casimir invariants when $W_{1,1}=0$. If $W_{1,1}\not= 0$, the Poisson matrix whose determinant is $16W_{1,1}^2$ is invertible and the corresponding Poisson bracket has no Casimir invariants. As a consequence, if $W_{1,1}=0$ there are only two effective dynamical variables, namely $(P_{1,0},P_{0,1})$, whereas there are four effective dynamical variables otherwise.  The resulting dynamical system has one macroscopic degree of freedom $(P_{1,0},P_{0,1})$ and one microscopic one $(W_{2,0},W_{0,2})$, provided $W_{1,1}\not= 0$. It is worth noticing that the microscopic degree of freedom, describing the shape of the beam, is different than the one in Subalgebra 3 or 4. This time, the conjugate variables (although non-canonical) are $W_{2,0}=P_{2,0}-P_{1,0}^2$ and $W_{0,2}=P_{0,2}-P_{0,1}^2$. 

In terms of the moments $P$, the variables are $(P_{1,0},P_{0,1},P_{2,0},P_{0,2})$ and the corresponding bracket is 
\begin{eqnarray*}
\{F,G\}&=&\frac{\partial F}{\partial P_{1,0}}\frac{\partial G}{\partial P_{0,1}}-\frac{\partial F}{\partial P_{0,1}}\frac{\partial G}{\partial P_{1,0}}+2P_{1,0}\left( \frac{\partial F}{\partial P_{2,0}}\frac{\partial G}{\partial P_{0,1}}-\frac{\partial F}{\partial P_{0,1}}\frac{\partial G}{\partial P_{2,0}}\right)\\
&& +2P_{0,1}\left( \frac{\partial F}{\partial P_{1,0}}\frac{\partial G}{\partial P_{0,2}}-\frac{\partial F}{\partial P_{0,2}}\frac{\partial G}{\partial P_{1,0}}\right)+4P_{1,1}\left( \frac{\partial F}{\partial P_{2,0}}\frac{\partial G}{\partial P_{0,2}}-\frac{\partial F}{\partial P_{0,2}}\frac{\partial G}{\partial P_{2,0}},\right),
\end{eqnarray*}
where $P_{1,1}=P_{1,0}P_{0,1}+W_{1,1}(P_{2,0}-P_{1,0}^2,P_{0,2}-P_{0,1}^2)$, i.e., the off-diagonal term $W_{1,1}$ of the covariance matrix is an arbitrary function of the diagonal terms $W_{2,0}$ and $W_{0,2}$. This bracket is a Poisson bracket for any function $W_{1,1}$.  

\subsection{Hamiltonian closures for $(P_{1,0},P_{0,1},W_{0,2},W_{2,0},\ldots,W_{N,0})$}
\label{ex:1g}

We generalize the above example by  including higher order moments in position, i.e., by considering the dynamical variables $(P_{1,0},P_{0,1},W_{0,2},W_{2,0},\ldots,W_{N,0})$ for some $N\geq 2$, and the closure has to be performed on $W_{n,1}$ for all $n=1,\ldots, N-1$. Bracket~(\ref{eq:brackWT}) reads
\begin{equation}
\label{eq:PBex1}
\{F,G\}=\frac{\partial F}{\partial P_{1,0}}\frac{\partial G}{\partial P_{0,1}}-\frac{\partial F}{\partial P_{0,1}}\frac{\partial G}{\partial P_{1,0}}+2 (n+1) W_{n,1}\left( \frac{\partial F}{\partial W_{n+1,0}}\frac{\partial G}{\partial W_{0,2}}-\frac{\partial F}{\partial W_{0,2}}\frac{\partial G}{\partial W_{n+1,0}}\right), 
\end{equation}
with implicit summation from $n=1$ to $N-1$. The Jacobi identity~(\ref{eq:JacA}) reduces to the following conditions on the closures:
$$
W_{n,1}\frac{\partial W_{n',1}}{\partial W_{0,2}}-W_{n',1}\frac{\partial W_{n,1}}{\partial W_{0,2}}=0,
$$
for all $n,n'=1,\ldots, N-1$, which is easily solved as:
$$
W_{n,1}=\psi(W_{0,2},W_{2,0},\ldots,W_{N,0}) \phi_n(W_{2,0},\ldots,W_{N,0}),
$$
with $N-1$ arbitrary functions $\phi_n$ of $N-1$ variables (actually, one of these functions can be absorbed in a redefinition of $\psi$) and an arbitrary function $\psi$ of $N$ variables. 

The Casimir invariants $C$ of the Poisson bracket~(\ref{eq:PBex1}) satisfy the following equation
$$
\sum_{n=2}^{N} n\phi_{n-1}(W_{2,0},\ldots,W_{N,0})\frac{\partial C}{\partial W_{n,0}}=0,
$$
which can be solved using the method of characteristics. It leads to the existence of $N-2$ Casimir invariants. For instance, if $\phi_n=W_{n+1,0}/W_{2,0}$ then the $N-2$ Casimir invariants are given by $C=W_{n+1,0}/W_{2,0}^{(n+1)/2}$, and the closures write
\begin{equation}
\label{eq:3gWn1}
W_{n,1}=\frac{W_{1,1}W_{n+1,0}}{W_{2,0}},
\end{equation}
for $n=2,\ldots,N-1$. 

For the set of dynamical variables considered in this section, the problem of finding the right closures leading to a Poisson bracket is not very constrained, but on a minus side, the dynamics is itself very constrained with the presence of $N-2$ Casimir invariants which drastically reduces the number of effective dynamical variables. As in the example in Sec.~\ref{ex:1}, there are four effective dynamical variables, independently of $N$, i.e., two degrees of freedom, one macroscopic $(P_{1,0},P_{0,1})$ and one microscopic degree of freedom, e.g., $(W_{2,0},W_{0,2})$. 

\subsection{Hamiltonian closures for $(P_{1,0},P_{0,1}, W_{1,1},W_{0,2},W_{2,0},W_{3,0})$}  
\label{ex:2}

In this case, the variables are $(P_{1,0},P_{0,1}, W_{1,1},W_{0,2},W_{2,0},W_{3,0})$ and the closure has to be performed on $W_{2,1}$. The bracket is given by
\begin{eqnarray*}
\{F,G\}&=&\frac{\partial F}{\partial P_{1,0}}\frac{\partial G}{\partial P_{0,1}}-\frac{\partial F}{\partial P_{0,1}}\frac{\partial G}{\partial P_{1,0}}+2W_{0,2}\left( \frac{\partial F}{\partial W_{1,1}}\frac{\partial G}{\partial W_{0,2}}-\frac{\partial F}{\partial W_{0,2}}\frac{\partial G}{\partial W_{1,1}}\right)\\
&& -2W_{2,0}\left( \frac{\partial F}{\partial W_{1,1}}\frac{\partial G}{\partial W_{2,0}}-\frac{\partial F}{\partial W_{2,0}}\frac{\partial G}{\partial W_{1,1}}\right)+4W_{1,1}\left( \frac{\partial F}{\partial W_{2,0}}\frac{\partial G}{\partial W_{0,2}}-\frac{\partial F}{\partial W_{0,2}}\frac{\partial G}{\partial W_{2,0}}\right)\\
&& -3W_{3,0}\left( \frac{\partial F}{\partial W_{1,1}}\frac{\partial G}{\partial W_{3,0}}-\frac{\partial F}{\partial W_{3,0}}\frac{\partial G}{\partial W_{1,1}}\right)-6W_{2,1}\left( \frac{\partial F}{\partial W_{0,2}}\frac{\partial G}{\partial W_{3,0}}-\frac{\partial F}{\partial W_{3,0}}\frac{\partial G}{\partial W_{0,2}}\right),
\end{eqnarray*}
where $W_{2,1}=\sqrt{W_{2,0}}\widetilde{W}_{2,1}(W_{1,1},W_{2,0}W_{0,2},W_{3,0}/W_{2,0}^{3/2})$ since $W_{1,1}$ is among the variables. In order to have a Poisson bracket, the following equation on $\widetilde{W}_{2,1}(W_{1,1},X_{0,2},X_{3,0})$, where $(X_{0,2},X_{3,0})=(W_{2,0}W_{0,2},W_{3,0}/W_{2,0}^{3/2})$, has to be satisfied:
$$
\frac{\partial \widetilde{W}_{2,1}}{\partial W_{1,1}}+2W_{1,1}\frac{\partial \widetilde{W}_{2,1}}{\partial X_{0,2}}=X_{3,0},
$$ 
whose solution is given by
$$
\widetilde{W}_{2,1}(W_{1,1},X_{0,2},X_{3,0})=W_{1,1}X_{3,0}+\overline{W}_{2,1}(X_{0,2}-W_{1,1}^2,X_{3,0}),
$$
with an arbitrary function $\overline{W}_{2,1}$. In other words, using the non-rescaled variables, the closure $W_{2,1}$ is written as
\begin{equation}
\label{eq:C2}
W_{2,1}=\frac{W_{1,1}W_{3,0}}{W_{2,0}}+\sqrt{W_{2,0}}\, \overline{W}_{2,1}(W_{2,0}W_{0,2}-W_{1,1}^2,W_{3,0}/W_{2,0}^{3/2}).
\end{equation}
First we notice that contrary to the above examples, the closure $W_{2,1}=0$ does not lead to a Hamiltonian system. This departs from the finite-dimensional truncation reexamined in Sec.~\ref{sec:FDLP}. 

It should be noticed that when $\overline{W}_{2,1}=0$, the corresponding Poisson bracket has two Casimir invariants, $C_1=W_{0,2}W_{2,0}-W_{1,1}^2$ and $C_2=W_{3,0}/W_{2,0}^{3/2}$. Therefore, as in the example of Sec.~\ref{ex:1}, the arbitrary function $\overline{W}_{2,1}$ depends on the Casimir invariants of the Poisson bracket when $\overline{W}_{2,1}=0$. We also notice that the solution resembles Eq.~(\ref{eq:3gWn1}), even if this time, $W_{1,1}$ is among the dynamical variables. 

For $\overline{W}_{2,1}\not= 0$, there is no Casimir invariant since the determinant of the Poisson matrix is equal to $144W_{2,0}^3\overline{W}_{2,1}^2$.
Therefore we conclude that there are four effective dynamical variables (i.e., two degrees of freedom) if $\overline{W}_{2,1}=0$ and six effective variables (i.e., three degrees of freedom) otherwise. 

This closure is interesting because this subalgebra includes the kinetic energy term $P_{0,2}/2$ and allows cubic potentials (i.e., quadratic forces), and it contains two microscopic degree of freedom for the dynamics of the shape of the beam. Expressed in the dynamical variables $(Y_{0,2},X_{3,0})=(W_{0,2}W_{2,0}-W_{1,1}^2,W_{3,0}/W_{2,0}^{3/2})$, the Poisson bracket writes
\begin{eqnarray*}
\{F,G\}&=& \frac{\partial F}{\partial P_{1,0}}\frac{\partial G}{\partial P_{0,1}}-\frac{\partial F}{\partial P_{0,1}}\frac{\partial G}{\partial P_{1,0}}+2W_{2,0}\left(\frac{\partial F}{\partial W_{2,0}}\frac{\partial G}{\partial W_{1,1}}-\frac{\partial F}{\partial W_{1,1}}\frac{\partial G}{\partial W_{2,0}} \right)\\
&& \qquad +6\overline{W}_{2,1}(Y_{0,2},X_{3,0})\left(\frac{\partial F}{\partial X_{3,0}}\frac{\partial G}{\partial Y_{0,2}}-\frac{\partial F}{\partial Y_{0,2}}\frac{\partial G}{\partial X_{3,0}} \right),
\end{eqnarray*} 
which means that all the three degrees of freedom are somehow decoupled, i.e., the two microscopic degrees of freedom are described by the two pairs of conjugate variables $(W_{2,0},W_{1,1})$ (in a similar way as in Subalgebras 3 and 4) and $(Y_{3,0},X_{0,2})$. 

\subsection{Hamiltonian closures for $(P_{1,0},P_{0,1}, W_{1,1},W_{0,2},W_{2,0},\ldots,W_{N,0})$}
\label{ex:2g}

We generalize the above example by including higher order moments in the position. We consider the dynamical variables $(P_{1,0},P_{0,1}, W_{1,1},W_{0,2},W_{2,0},\ldots,W_{N,0})$ for $N\geq 3$ and the closure has to be performed on $W_{n,1}$ for $n=2,\ldots, N-1$. Bracket~(\ref{eq:brackWT}) reduces to 
\begin{eqnarray*}
\{F,G\}&=&\frac{\partial F}{\partial P_{1,0}}\frac{\partial G}{\partial P_{0,1}}-\frac{\partial F}{\partial P_{0,1}}\frac{\partial G}{\partial P_{1,0}}+2W_{0,2}\left( \frac{\partial F}{\partial W_{1,1}}\frac{\partial G}{\partial W_{0,2}}-\frac{\partial F}{\partial W_{0,2}}\frac{\partial G}{\partial W_{1,1}}\right)\\
&& +nW_{n,0}\left( \frac{\partial F}{\partial W_{n,0}}\frac{\partial G}{\partial W_{1,1}}-\frac{\partial F}{\partial W_{1,1}}\frac{\partial G}{\partial W_{n,0}}\right)+2nW_{n-1,1}\left( \frac{\partial F}{\partial W_{n,0}}\frac{\partial G}{\partial W_{0,2}}-\frac{\partial F}{\partial W_{0,2}}\frac{\partial G}{\partial W_{n,0}}\right),
\end{eqnarray*}
with implicit summation from $n=2$ to $N$. We notice that the Poisson bracket is linear in the variables as in the cases examined above. Since $W_{1,1}$ is among the dynamical variables, the closures are written as
$$
W_{n,1}=W_{2,0}^{(n-1)/2}\widetilde{W}_{n,1}\left( W_{1,1},W_{0,2}W_{2,0},W_{3,0}/W_{2,0}^{3/2},\ldots,
W_{N,0}/W_{2,0}^{N/2} \right). 
$$
The Jacobi identity reduces to the following constraints on the closures $\widetilde{W}_{n,1}(W_{1,1},X_{0,2},X_{3,0},\ldots,X_{N,0})$, where $(X_{0,2},X_{3,0},\ldots,X_{N,0})=(W_{0,2}W_{2,0},W_{3,0}/W_{2,0}^{3/2},\ldots,
W_{N,0}/W_{2,0}^{N/2})$:
\begin{equation}
 \label{eq:2aJac1}
X_{n,0}-\frac{\partial \widetilde{W}_{n-1,1}}{\partial W_{1,1}}-2W_{1,1}\frac{\partial \widetilde{W}_{n-1,1}}{\partial X_{0,2}}=0,
\end{equation}
for $n=3,\ldots, N$, and
\begin{equation}
\label{eq:2aJac2}
X_{n,0}\frac{\partial \widetilde{W}_{k-1,1}}{\partial W_{1,1}}-X_{k,0}\frac{\partial \widetilde{W}_{n-1,1}}{\partial W_{1,1}}+2\widetilde{W}_{n-1,1}\frac{\partial \widetilde{W}_{k-1,1}}{\partial X_{0,2}}-2\widetilde{W}_{k-1,1}\frac{\partial \widetilde{W}_{n-1,1}}{\partial X_{0,2}}=0,
\end{equation} 
for $n=4,\ldots, N$ and $k=3,\ldots, n-1$. The solution of Eq.~(\ref{eq:2aJac1}) writes
$$
\widetilde{W}_{n-1,1}=W_{1,1}X_{n,0}+\overline{W}_{n-1,1}\left(X_{0,2}-W_{1,1}^2,X_{3,0},\ldots,X_{N,0} \right). 
$$
Inserting these expressions for the closures into Eq.~(\ref{eq:2aJac2}) leads to 
$$
\overline{W}_{n-1,1}\frac{\partial \overline{W}_{k-1,1}}{\partial Y_{0,2}}-\overline{W}_{k-1,1}\frac{\partial \overline{W}_{n-1,1}}{\partial Y_{0,2}}=0,
$$
where $Y_{0,2}=X_{0,2}-W_{1,1}^2$.
The solution is given by
$$
\overline{W}_{n,1}\left(Y_{0,2},X_{3,0},\ldots,X_{N,0} \right)=\psi\left(Y_{0,2},X_{3,0},\ldots,W_{X,0} \right) \phi_n\left(X_{3,0},\ldots,X_{N,0} \right),
$$
for $n=2,\ldots,N-1$, and for arbitrary functions $\phi_n$ and $\psi$.
In summary, the closures are given by
\begin{eqnarray*}
W_{n,1}&=&\frac{W_{1,1}W_{n+1,0}}{W_{2,0}}+W_{2,0}^{(n-1)/2}\psi\left(W_{0,2}W_{2,0}-W_{1,1}^2,W_{3,0}/W_{2,0}^{3/2},\ldots,W_{N,0}/W_{2,0}^{N/2} \right)\\
&& \qquad \qquad \qquad \qquad\qquad \qquad\times \phi_n\left(W_{3,0}/W_{2,0}^{3/2},\ldots,W_{N,0}/W_{2,0}^{N/2} \right),
\end{eqnarray*}
for $n=2,\ldots, N-1$, and where $\psi$ and $\phi_n$ for $n=2,\ldots, N-1$ are arbitrary functions (actually, we can choose one of these functions to be equal to one, absorbing it in the definition of $\psi$). As in Sec.~\ref{ex:2}, we notice that the closure $W_{n,1}=0$ is not a Hamiltonian closure. 

For $\psi=0$, there are $N-1$ Casimir invariants, $W_{0,2}W_{2,0}-W_{1,1}^2$ and $W_{n,0}/W_{2,0}^{n/2}$ for $n=3,\ldots,N$. It should be noticed that these are Casimir invariants only in the case where the closures satisfy $W_{n,1}=W_{1,1}W_{n+1,0}/W_{2,0}$, which is also the expression of the closure in the example of Sec.~\ref{ex:1g} [see Eq.~(\ref{eq:3gWn1})].  
 
If $\psi\not= 0$, the resulting Poisson bracket has $N-3$ Casimir invariants of the form $C(W_{3,0}/W_{2,0}^{3/2},\dots,W_{N,0}/W_{2,0}^{N/2})$ where $C(X_{3,0}\ldots,X_{N,0})$ satisfies
$$
\sum_{n=3}^N n\phi_{n-1}(X_{3,0}\ldots,X_{N,0})\frac{\partial C}{\partial X_{n+1,0}}=0,
$$
which can be solved using the method of characteristics.
Therefore, if $\psi=0$, there are four effective dynamical variables whereas there are six of these effective variables if $\psi\not= 0$, in a very analogous way as in the example of Sec.~\ref{ex:2}. Adding more momenta in the position does not necessarily lead to an increase of dynamical complexity, at least as far as the Poisson bracket is concerned. 

\subsection{Hamiltonian closures for $(P_{1,0},P_{0,1}, W_{1,1},W_{2,1},W_{0,2},W_{2,0},W_{3,0})$}
\label{ex:3}

So far, the truncated brackets we have examined were linear in the variables and the closures, whereas, in the general case, the coefficients in Bracket~(\ref{eq:brackWT}) are quadratic. By including $W_{2,1}$ as dynamical variable, the truncated bracket given by Eq.~(\ref{eq:brackWT}) contains at least one quadratic term. First we consider the case with only one quadratic term. The set of variables is $(P_{1,0},P_{0,1}, W_{1,1},W_{2,1},W_{0,2},W_{2,0},W_{3,0})$. The closures are for $W_{1,2}$ and $W_{4,0}$. There is a term in the bracket of the form 
$$
3(W_{2,0}^2-W_{4,0})\frac{\partial F}{\partial W_{2,1}}\frac{\partial G}{\partial W_{3,0}}.
$$ 
Since $W_{1,1}$ is among the variables, the closures are constrained by
\begin{eqnarray*}
&& W_{4,0}=W_{2,0}^{2}\widetilde{W}_{4,0}\left( W_{1,1},W_{2,1}/W_{2,0}^{1/2},W_{0,2}W_{2,0},W_{3,0}/W_{2,0}^{3/2} \right),\\
&& W_{1,2}=W_{2,0}^{-1/2}\widetilde{W}_{1,2}\left( W_{1,1},W_{2,1}/W_{2,0}^{1/2},W_{0,2}W_{2,0},W_{3,0}/W_{2,0}^{3/2} \right).
\end{eqnarray*}
The Jacobi identity reduces to the following three coupled partial differential equations on $\widetilde{W}_{4,0}(W_{1,1},X_{2,1},X_{0,2},X_{3,0})$ and $\widetilde{W}_{1,2}(W_{1,1},X_{2,1},X_{0,2},X_{3,0})$:
\begin{eqnarray}
&& 2 X_{2,1}-X_{3,0}\frac{\partial \widetilde{W}_{1,2}}{\partial X_{2,1}}-\frac{\partial \widetilde{W}_{1,2}}{\partial W_{1,1}}-2W_{1,1}\frac{\partial \widetilde{W}_{1,2}}{\partial X_{0,2}}=0,\label{eq:ex3_J1}\\
&& X_{3,0}\frac{\partial \widetilde{W}_{4,0}}{\partial X_{2,1}}+\frac{\partial \widetilde{W}_{4,0}}{\partial W_{1,1}}+2W_{1,1}\frac{\partial \widetilde{W}_{4,0}}{\partial X_{0,2}}=0,\label{eq:ex3_J2}\\
&& 3(X_{2,1}-X_{3,0}W_{1,1})\frac{\partial \widetilde{W}_{4,0}}{\partial X_{3,0}}+X_{0,2}\frac{\partial \widetilde{W}_{4,0}}{\partial W_{1,1}}+(2\widetilde{W}_{1,2}-X_{2,1}W_{1,1})\frac{\partial \widetilde{W}_{4,0}}{\partial X_{2,1}}+2X_{0,2}W_{1,1}\frac{\partial \widetilde{W}_{4,0}}{\partial X_{0,2}}+4W_{1,1}(\widetilde{W}_{4,0}-1)\nonumber \\
&& \qquad \qquad \qquad =2(\widetilde{W}_{4,0}-1)\frac{\partial \widetilde{W}_{1,2}}{\partial X_{2,1}}+2X_{3,0}\frac{\partial \widetilde{W}_{1,2}}{\partial W_{1,1}}+4X_{2,1}\frac{\partial \widetilde{W}_{1,2}}{\partial X_{0,2}}. \label{eq:ex3_J3}
\end{eqnarray}
First we notice that $\widetilde{W}_{1,2}=0$ or $\widetilde{W}_{4,0}=0$ are not solutions of the above equations, so setting the closure functions to zero does not correspond to a Hamiltonian closure, unlike the finite-dimensional Lie-Poisson algrebas of Sec.~\ref{sec:FDLP}.
 
The first two equations have solutions of the form:
\begin{eqnarray*}
&& \widetilde{W}_{1,2}=2X_{2,1}W_{1,1}-W_{1,1}^2X_{3,0}+\overline{W}_{1,2}\left(X_{0,2}-W_{1,1}^2,X_{2,1}-X_{3,0}W_{1,1},X_{3,0} \right),\\ 
&& \widetilde{W}_{4,0}=1+X_{3,0}^2+\overline{W}_{4,0}\left(X_{0,2}-W_{1,1}^2,X_{2,1}-X_{3,0}W_{1,1},X_{3,0} \right).
\end{eqnarray*}
Using the change of variables $(Y_{0,2},Y_{2,1})=(X_{0,2}-W_{1,1}^2,X_{2,1}-X_{3,0}W_{1,1})$, the third equation becomes
$$
 2\overline{W}_{4,0}\frac{\partial \overline{W}_{1,2}}{\partial Y_{2,1}}-2\overline{W}_{1,2}\frac{\partial \overline{W}_{4,0}}{\partial Y_{2,1}} +Y_{0,2}X_{3,0}\frac{\partial \overline{W}_{4,0}}{\partial Y_{2,1}}-3Y_{2,1}\frac{\partial \overline{W}_{4,0}}{\partial X_{3,0}}+4Y_{2,1}\frac{\partial \overline{W}_{1,2}}{\partial Y_{0,2}}=2Y_{2,1}X_{3,0},
$$
for $\overline{W}_{4,0}(Y_{0,2},Y_{2,1},X_{3,0})$ and $\overline{W}_{1,2}(Y_{0,2},Y_{2,1},X_{3,0})$. 
We notice that a particular solution of the above equation is given by
\begin{eqnarray}
&& \overline{W}^{(0)}_{4,0}=0,\label{eq:partsol40}\\ 
&& \overline{W}^{(0)}_{1,2}=\frac{1}{2}Y_{0,2}X_{3,0}. \label{eq:partsol12}
\end{eqnarray}
Next we perform the change of functions $ \overline{W}_{1,2}=\overline{W}^{(0)}_{1,2}+\overline{\overline{W}}_{1,2}$.
The equation on the closures becomes
$$
\overline{\overline{W}}_{1,2}\frac{\partial \overline{W}_{4,0}}{\partial Y_{2,1}}-\overline{W}_{4,0}\frac{\partial \overline{\overline{W}}_{1,2}}{\partial Y_{2,1}}
+\frac{3}{2}Y_{2,1}\frac{\partial \overline{W}_{4,0}}{\partial X_{3,0}}-2Y_{2,1}\frac{\partial \overline{\overline{W}}_{1,2}}{\partial Y_{0,2}}=0. 
$$
This equation is linear in each of the closure functions $\overline{\overline{W}}_{1,2}$ and $\overline{W}_{4,0}$, so it can be solved using the method of characteristics. This equation links both closures and there is an infinite set of solutions. These solutions are given by
\begin{eqnarray*}
&& \overline{\overline{W}}_{1,2}=-\frac{3}{2}Y_{2,1}\frac{\partial \overline{C}/\partial X_{3,0}}{\partial \overline{C}/\partial Y_{2,1}},\\
&& \overline{W}_{4,0}=-2 Y_{2,1}\frac{\partial \overline{C}/\partial Y_{0,2}}{\partial \overline{C}/\partial Y_{2,1}}, 
\end{eqnarray*}
where $\overline{C}=\overline{C}(Y_{0,2},Y_{2,1},X_{3,0})$ is an arbitrary function with non-zero derivative with respect to $Y_{2,1}$. In summary, all the Hamiltonian closures are given by
\begin{eqnarray}
&& W_{1,2}=\frac{W_{2,1}W_{1,1}+W_{0,2}W_{3,0}}{2W_{2,0}}-\frac{3}{2}\frac{W_{2,1}W_{2,0}-W_{1,1}W_{3,0}}{W_{2,0}}\frac{\partial C/\partial W_{3,0}}{\partial C/\partial W_{2,1}}, \label{eq:ex3W12} \\
&& W_{4,0}=W_{2,0}^2+\frac{W_{3,0}^2}{W_{2,0}}-2\frac{W_{2,1}W_{2,0}-W_{1,1}W_{3,0}}{W_{2,0}}\frac{\partial C/\partial W_{0,2}}{\partial C/\partial W_{2,1}},\label{eq:ex3W40}
\end{eqnarray}
for 
\begin{equation}
\label{eq:ex3C}
C(W_{1,1},W_{2,1},W_{0,2},W_{2,0},W_{3,0})=\overline{C}\left(W_{0,2}W_{2,0}-W_{1,1}^2,\frac{W_{2,1}W_{2,0}-W_{1,1}W_{3,0}}{W_{2,0}^{3/2}},\frac{W_{3,0}}{W_{2,0}^{3/2}} \right).
\end{equation}   
It turns out that $C$ given by Eq.~(\ref{eq:ex3C}) is a Casimir invariant of the Hamiltonian system defined by the closure functions given by Eqs.~(\ref{eq:ex3W12})-(\ref{eq:ex3W40}), and it is the only one. In other words, there is a one to one correspondance between the closure functions and the Casimir invariant of the system. 

Like the cases in Secs.~\ref{ex:2} and \ref{ex:2g}, the resulting Hamiltonian system possesses six effective dynamical variables, one macroscopic degree of freedom and two degrees of freedom for the shape of the beam. 

{\em Remark}: Since the number of variables is odd, the resulting truncated Poisson bracket necessarily has at least one Casimir invariant. In this case, it possesses just one Casimir invariant. This Casimir invariant reduces the effective dimension of phase space. By inverting Eq.~(\ref{eq:ex3C}), the constraint on the dynamics imposed by this invariant can be rewritten as
$$
\frac{W_{2,1}W_{2,0}-W_{1,1}W_{3,0}}{W_{2,0}^{3/2}}=\gamma\left(W_{0,2}W_{2,0}-W_{1,1}^2,\frac{W_{3,0}}{W_{2,0}^{3/2}} \right).
$$ 
This corresponds, for instance, to a constraint on the dynamical variable $W_{2,1}$. So if we drop $W_{2,1}$ as dynamical variable, we end up in the case of Sec.~\ref{ex:2}. The above equation corresponds to the closure obtained in Sec.~\ref{ex:2}, and in particular to Eq.~(\ref{eq:C2}). We notice that if $W_{2,1}$ ceases to be a dynamical variable, the closure functions $W_{1,2}$ and $W_{4,0}$ do no longer appear in the truncated bracket since the partial derivatives with respect to $W_{2,1}$ vanish, and therefore these closures can be discarded.  

\subsection{Hamiltonian closures for $(P_{1,0},P_{0,1}, W_{1,1},W_{2,1},W_{0,2},W_{2,0},W_{3,0},W_{4,0})$}
\label{ex:3b}

We consider the set of dynamical variables $(P_{1,0},P_{0,1}, W_{1,1},W_{2,1},W_{0,2},W_{2,0},W_{3,0},W_{4,0})$. This example generalizes the above example by including $W_{4,0}$ as a dynamical variables. The closure functions are now $W_{1,2}$, $W_{3,1}$ and $W_{5,0}$. As in the example of Sec.~\ref{ex:3} the Poisson bracket contains terms which are quadratic in the dynamical variables, namely
$$
n(W_{2,0}W_{n-1,0}-W_{n+1,0})\left( \frac{\partial F}{\partial W_{2,1}}\frac{\partial G}{\partial W_{n,0}}-\frac{\partial F}{\partial W_{n,0}}\frac{\partial G}{\partial W_{2,1}}\right),
$$
for $n=3,4$. Since $W_{1,1}$ is a dynamical variables, the closure can be written as
\begin{eqnarray*}
&& W_{3,1}=W_{2,0}\widetilde{W}_{3,1}\left( W_{1,1},W_{2,1}/W_{2,0}^{1/2},W_{0,2}W_{2,0},W_{3,0}/W_{2,0}^{3/2},
W_{4,0}/W_{2,0}^{2} \right),\\
&& W_{5,0}=W_{2,0}^{5/2}\widetilde{W}_{5,0}\left( W_{1,1},W_{2,1}/W_{2,0}^{1/2},W_{0,2}W_{2,0},W_{3,0}/W_{2,0}^{3/2},W_{4,0}/W_{2,0}^{2} \right),\\
&& W_{1,2}=W_{2,0}^{-1/2}\widetilde{W}_{1,2}\left( W_{1,1},W_{2,1}/W_{2,0}^{1/2},W_{0,2}W_{2,0},W_{3,0}/W_{2,0}^{3/2},W_{4,0}/W_{2,0}^{2} \right).
\end{eqnarray*}
For $\widetilde{W}_{5,0}(W_{1,1},X_{2,1},X_{0,2},X_{3,0},X_{4,0})$ we have the following two equations:
\begin{eqnarray*}
&& X_{3,0}\frac{\partial \widetilde{W}_{5,0}}{\partial X_{2,1}}+\frac{\partial \widetilde{W}_{5,0}}{\partial W_{1,1}}+2W_{1,1}\frac{\partial \widetilde{W}_{5,0}}{\partial X_{0,2}}=0,\\
&& (X_{4,0}-1)\frac{\partial \widetilde{W}_{5,0}}{\partial X_{2,1}}+X_{3,0}\frac{\partial \widetilde{W}_{5,0}}{\partial W_{1,1}}+2X_{2,1}\frac{\partial \widetilde{W}_{5,0}}{\partial X_{0,2}}=0,
\end{eqnarray*}
which are solved as
$$
\widetilde{W}_{5,0}=\overline{W}_{5,0}\left( (X_{0,2}-W_{1,1}^2)(X_{4,0}-1-X_{3,0}^2)-(X_{2,1}-W_{1,1}X_{3,0})^2,X_{3,0},X_{4,0}\right),
$$
where $\overline{W}_{5,0}$ is an arbitrary function of three variables. For $\widetilde{W}_{3,1}$ we have the following two equations:
\begin{eqnarray*}
&& X_{3,0}\frac{\partial \widetilde{W}_{3,1}}{\partial X_{2,1}}+\frac{\partial \widetilde{W}_{3,1}}{\partial W_{1,1}}+2W_{1,1}\frac{\partial \widetilde{W}_{3,1}}{\partial X_{0,2}}=X_{4,0},\\
&&(X_{4,0}-1) \frac{\partial \widetilde{W}_{3,1}}{\partial X_{2,1}}+X_{3,0}\frac{\partial \widetilde{W}_{3,1}}{\partial W_{1,1}}+2X_{2,1}\frac{\partial \widetilde{W}_{3,1}}{\partial X_{0,2}}=\widetilde{W}_{5,0}-X_{3,0},
\end{eqnarray*}
which is solved as 
\begin{eqnarray*}
\widetilde{W}_{3,1}&=&W_{1,1}X_{4,0}+\frac{X_{2,1}-W_{1,1}X_{3,0}}{X_{4,0}-1-X_{3,0}^2} (\overline{W}_{5,0}-X_{3,0}-X_{3,0}X_{4,0})\\
&& \quad +\overline{W}_{3,1}\left((X_{0,2}-W_{1,1}^2)(X_{4,0}-1-X_{3,0}^2)-(X_{2,1}-W_{1,1}X_{3,0})^2 ,X_{3,0},X_{4,0}\right),
\end{eqnarray*}
where $\overline{W}_{3,1}$ is an arbitrary function. For $\widetilde{W}_{1,2}$ we have the two equations:
\begin{eqnarray*}
&& X_{3,0}\frac{\partial \widetilde{W}_{1,2}}{\partial X_{2,1}}+\frac{\partial \widetilde{W}_{1,2}}{\partial W_{1,1}}+2W_{1,1}\frac{\partial \widetilde{W}_{1,2}}{\partial X_{0,2}}=2X_{2,1},\\
&&(X_{4,0}-1) \frac{\partial \widetilde{W}_{1,2}}{\partial X_{2,1}}+X_{3,0}\frac{\partial \widetilde{W}_{1,2}}{\partial W_{1,1}}+2X_{2,1}\frac{\partial \widetilde{W}_{1,2}}{\partial X_{0,2}}=2(\widetilde{W}_{3,1}-W_{1,1}),
\end{eqnarray*}
which have the following solutions:
\begin{eqnarray*}
\widetilde{W}_{1,2}&=&2W_{1,1} X_{2,1}-W_{1,1}^2X_{3,0}+\left(\frac{X_{2,1}-W_{1,1}X_{3,0}}{X_{4,0}-1-X_{3,0}^2}\right)^2\left(\overline{W}_{5,0}-2X_{3,0}X_{4,0}+X_{3,0}^3\right)\\
&& \quad +2\frac{X_{2,1}-W_{1,1}X_{3,0}}{X_{4,0}-1-X_{3,0}^2}\overline{W}_{3,1}+\overline{W}_{1,2},
\end{eqnarray*}
where all the functions $\overline{W}_{5,0}$, $\overline{W}_{3,1}$ and $\overline{W}_{1,2}$ are functions of $X=(X_{0,2}-W_{1,1}^2)(X_{4,0}-1-X_{3,0}^2)-(X_{2,1}-W_{1,1}X_{3,0})^2$, $X_{3,0}$ and $X_{4,0}$.
We notice that the closures are in general singular since $Y=X_{4,0}-1-X_{3,0}^2$ is present in the denominator. In order to obtain non-singular closures, the closure functions are of the following form:
\begin{eqnarray*}
&& \overline{W}_{5,0}=2X_{3,0}(Y+1)+X_{3,0}^3+Y^2\overline{\overline{W}}_{5,0}\left(X,X_{3,0},Y\right),\\
&& \overline{W}_{3,1}=Y \overline{\overline{W}}_{3,1}\left(X,X_{3,0},Y\right).
\end{eqnarray*} 
These three functions are linked by one last remaining equation which is
\begin{eqnarray*}
&& 4 Y^2\left(\overline{\overline{W}}_{3,1}\frac{\partial \overline{\overline{W}}_{5,0}}{\partial Y}-\overline{\overline{W}}_{5,0}\frac{\partial\overline{\overline{W}}_{3,1} }{\partial Y}\right)+4Y\left(\overline{W}_{1,2}\frac{\partial \overline{\overline{W}}_{3,1}}{\partial X}-\overline{\overline{W}}_{3,1}\frac{\partial \overline{W}_{1,2} }{\partial X}\right) \\
&& \qquad +4XY \left(\overline{\overline{W}}_{3,1}\frac{\partial \overline{\overline{W}}_{5,0}}{\partial X}-\overline{\overline{W}}_{5,0}\frac{\partial\overline{\overline{W}}_{3,1} }{\partial X} \right) +2X_{3,0}Y\frac{\partial \overline{\overline{W}}_{3,1}}{\partial Y}-3Y\frac{\partial \overline{\overline{W}}_{3,1}}{\partial X_{3,0}}+4X_{3,0}\overline{\overline{W}}_{3,1}=0.
\end{eqnarray*}
For $Y=0$, we notice that a necessary condition is that $\overline{\overline{W}}_{3,1}(X,X_{3,0},0)=0$ for all $X$ and $X_{3,0}$, if we assume that all the first order derivatives of the closure functions are finite at $Y=0$. By differentiating the above equation with respect to $Y$, we show that a necessary condition is that $\partial \overline{\overline{W}}_{3,1}/\partial Y=0$ at $Y=0$. By iteration, we show that all higher order derivatives of $\overline{\overline{W}}_{3,1}$ with respect to $Y$ vanish at $Y=0$. If we assume that $\overline{\overline{W}}_{3,1}$ is analytic in $Y$, then the only solution is $\overline{\overline{W}}_{3,1}=0$.     
In this case, the Jacobi identity is automatically satisfied for all functions $\overline{W}_{1,2}$ and $\overline{\overline{W}}_{5,0}$. 

In summary, the non-singular Hamiltonian closures are given by
\begin{eqnarray*}
&& W_{1,2}=W_{1,1}\left(\frac{2W_{2,1}W_{2,0}-W_{1,1}W_{3,0}}{W_{2,0}^2}\right)+(W_{2,1}W_{2,0}-W_{1,1}W_{3,0})^2)W_{2,0}^{-7/2}\overline{\overline{W}}_{5,0}+\overline{W}_{1,2},\\
&& \widetilde{W}_{3,1}=\frac{W_{1,1}W_{4,0}W_{2,0}+W_{3,0}(W_{2,1}W_{2,0}-W_{1,1}W_{3,0})}{W_{2,0}^2}+(W_{2,1}W_{2,0}-W_{1,1}W_{3,0})(W_{4,0}W_{2,0}-W_{2,0}^3-W_{3,0}^2)W_{2,0}^{-7/2}\overline{\overline{W}}_{5,0},\\
&& \widetilde{W}_{5,0}=2\frac{W_{3,0}W_{4,0}}{W_{2,0}}-\frac{W_{3,0}^3}{W_{2,0}^2}+(W_{4,0}W_{2,0}-W_{2,0}^3-W_{3,0}^2)^2W_{2,0}^{-7/2}\overline{\overline{W}}_{5,0},
\end{eqnarray*}
for arbitrary functions $\overline{W}_{1,2}$ and $\overline{\overline{W}}_{5,0}$ of the three variables $X=[(W_{0,2}W_{2,0}-W_{1,1}^2)(W_{4,0}W_{2,0}-W_{2,0}^3-W_{3,0}^2)-(W_{2,1}W_{2,0}-W_{1,1}W_{3,0})^2]/W_{2,0}^3$, $X_{3,0}=W_{3,0}/W_{2,0}^{3/2}$ and $Y=(W_{4,0}W_{2,0}-W_{2,0}^3-W_{3,0}^2)/W_{2,0}^2$. 

The necessary and sufficient condition for the Poisson bracket to possess Casimir invariants of the form $\overline{C}(X,X_{3,0},Y)$ is 
\begin{equation}
4(X\overline{\overline{W}}_{5,0}-\overline{W}_{1,2})\frac{\partial \overline{C}}{\partial X}+3\frac{\partial \overline{C}}{\partial X_{3,0}}+(4Y\overline{\overline{W}}_{5,0}-2X_{3,0})\frac{\partial C}{\partial Y}=0, \\
\end{equation}
which can be solved using the method of characteristics. It leads to the existence of two Casimir invariants (whose expressions depend on the closure functions $\overline{\overline{W}}_{5,0}$ and $\overline{W}_{1,2}$). The presence of Casimir invariants reduces the number of effective dynamical variables. The number of effective dynamical variables is again six as in Secs.~\ref{ex:2},\ref{ex:2g} and \ref{ex:3}. Despite the fact that the number of variables is even, the increase of the dimension of phase space by adding moments in the position only does not lead to an increase of dynamical complexity.

\subsection{Hamiltonian closures for $(P_{1,0},P_{0,1}, W_{1,1},W_{2,1},W_{0,2},W_{2,0},\ldots,W_{N,0})$}
\label{ex:3g}

Here we consider the set of variables $(P_{1,0},P_{0,1}, W_{1,1},W_{2,1},W_{0,2},W_{2,0},\ldots,W_{N,0})$ for $N\geq 5$ which generalizes the above examples by including additional moments in the position. The closures have to be performed on $W_{n,1}$ for $n=3,\ldots, N-1$ as it can be expected from the example in Sec.~\ref{ex:2g}, and in addition on $W_{1,2}$ and $W_{N+1,0}$ which is similar to the cases in Secs.~\ref{ex:3} and \ref{ex:3b}. As in the above cases with $W_{2,1}$ a dynamical variable, the Poisson bracket is not linear in $W$s since the following quadratic terms are introduced,
$$
n(W_{2,0}W_{n-1,0}-W_{n+1,0})\left( \frac{\partial F}{\partial W_{2,1}}\frac{\partial G}{\partial W_{n,0}}-\frac{\partial F}{\partial W_{n,0}}\frac{\partial G}{\partial W_{2,1}}\right),
$$
and one of these terms contains one of the closure functions, $W_{N+1,0}$. Since $W_{1,1}$ is among the dynamical variables, we write the closures as
\begin{eqnarray*}
&& W_{n,1}=W_{2,0}^{(n-1)/2}\widetilde{W}_{n,1}\left( W_{1,1},W_{2,1}/W_{2,0}^{1/2},W_{0,2}W_{2,0},W_{3,0}/W_{2,0}^{3/2},\ldots,
W_{N,0}/W_{2,0}^{N/2} \right),\\
&& W_{N+1,0}=W_{2,0}^{(N+1)/2}\widetilde{W}_{N+1,0}\left( W_{1,1},W_{2,1}/W_{2,0}^{1/2},W_{0,2}W_{2,0},W_{3,0}/W_{2,0}^{3/2},\ldots,
W_{N,0}/W_{2,0}^{N/2} \right),\\
&& W_{1,2}=W_{2,0}^{-1/2}\widetilde{W}_{1,2}\left( W_{1,1},W_{2,1}/W_{2,0}^{1/2},W_{0,2}W_{2,0},W_{3,0}/W_{2,0}^{3/2},\ldots,
W_{N,0}/W_{2,0}^{N/2} \right),
\end{eqnarray*}
for $n=3,\ldots, N-1$. Using the change of variables $(X_{2,1},X_{0,2},X_{3,0},\ldots,X_{N,0})= ( W_{2,1}/W_{2,0}^{1/2},W_{0,2}W_{2,0},W_{3,0}/W_{2,0}^{3/2},\ldots,
W_{N,0}/W_{2,0}^{N/2} )$, we 
first obtain some conditions on $W_{N+1,0}$:
\begin{eqnarray*}
&& X_{3,0}\frac{\partial \widetilde{W}_{N+1,0}}{\partial X_{2,1}}+\frac{\partial \widetilde{W}_{N+1,0}}{\partial W_{1,1}}+2W_{1,1}\frac{\partial \widetilde{W}_{N+1,0}}{\partial X_{0,2}}=0,\\
&& (X_{4,0}-1)\frac{\partial \widetilde{W}_{N+1,0}}{\partial X_{2,1}}+X_{3,0}\frac{\partial \widetilde{W}_{N+1,0}}{\partial W_{1,1}}+2X_{2,1}\frac{\partial \widetilde{W}_{N+1,0}}{\partial X_{0,2}}=0,\\
&& (X_{n+2,0}-X_{n,0})\frac{\partial \widetilde{W}_{N+1,0}}{\partial X_{2,1}}+X_{n+1,0}\frac{\partial \widetilde{W}_{N+1,0}}{\partial W_{1,1}}+2\widetilde{W}_{n,1}\frac{\partial \widetilde{W}_{N+1,0}}{\partial X_{0,2}}=0,
\end{eqnarray*}
for $n=3,\ldots,N-2$. Solving these $N-2$ coupled linear partial differential equations leads to 
\begin{equation}
\widetilde{W}_{N+1,0}=\widetilde{W}_{N+1,0}\left(X_{3,0},\ldots,X_{N,0} \right).
\label{eqn:ex3gWf0}
\end{equation}
Some conditions on $\widetilde{W}_{n,1}$ remain similar to the case of Sec.~\ref{ex:3b}:
\begin{eqnarray*}
&& X_{n+1,0}-X_{3,0}\frac{\partial \widetilde{W}_{n,1}}{\partial X_{2,1}}-\frac{\partial \widetilde{W}_{n,1}}{\partial W_{1,1}}-2W_{1,1}\frac{\partial \widetilde{W}_{n,1}}{\partial X_{0,2}}=0,\\
&& X_{n,0}-X_{n+2,0}+(X_{4,0}-1)\frac{\partial \widetilde{W}_{n,1}}{\partial X_{2,1}}+X_{3,0}\frac{\partial \widetilde{W}_{n,1}}{\partial W_{1,1}}+2X_{2,1}\frac{\partial \widetilde{W}_{n,1}}{\partial X_{0,2}}=0,
\end{eqnarray*}
for $n=3,\ldots,N-1$, which has a solution of the form
\begin{equation}
\label{eqn:ex3gWn1}
\widetilde{W}_{n,1}=X_{n+1,0}W_{1,1}+(X_{n+2,0}-X_{n,0}-X_{n+1,0}X_{3,0})\frac{X_{2,1}-W_{1,1}X_{3,0}}{X_{4,0}-1-X_{3,0}^2}+\overline{W}_{n,1}\left(X,X_{3,0},\ldots,X_{N,0} \right),
\end{equation}
where $X=(X_{4,0}-1-X_{3,0}^2)(X_{0,2}-X_{1,1}^2)-(X_{2,1}-W_{1,1}X_{3,0})^2$. 
Additional conditions on $\overline{W}_{n,1}$ are similar to some conditions obtained in Sec.~\ref{ex:2g}: 
$$
\overline{W}_{n,1}\frac{\partial \overline{W}_{n',1}}{\partial X}-\overline{W}_{n',1}\frac{\partial \overline{W}_{n,1}}{\partial X}=0,
$$
for all $n,n'=3,\ldots,N-1$, which are easily solved:
$$
\overline{W}_{n,1}=\psi(X,X_{3,0},\ldots,X_{N,0})\phi_n(X_{3,0},\ldots,X_{N,0}),
$$
for $n=3,\ldots,N-1$ and for arbitrary functions $\psi$ and $\phi_n$. 
 
Concerning the condition on $\widetilde{W}_{1,2}$, we have
\begin{eqnarray*}
&&2X_{2,1}-X_{3,0}\frac{\partial \widetilde{W}_{1,2}}{\partial X_{2,1}}-\frac{\partial \widetilde{W}_{1,2}}{\partial W_{1,1}}-2W_{1,1}\frac{\partial \widetilde{W}_{1,2}}{\partial X_{0,2}}=0,\\
&& 2(W_{1,1}-\widetilde{W}_{3,1})+(X_{4,0}-1)\frac{\partial \widetilde{W}_{1,2}}{\partial X_{2,1}}+X_{3,0}\frac{\partial \widetilde{W}_{1,2}}{\partial W_{1,1}}+2X_{2,1}\frac{\partial \widetilde{W}_{1,2}}{\partial X_{0,2}}=0.
\end{eqnarray*}
Together with the equation for $\overline{W}_{n,1}$, the general solution of the equations is
\begin{eqnarray}
\widetilde{W}_{1,2}&=&2W_{1,1}X_{2,1}-W_{1,1}^2X_{3,0}+(X_{5,0}-2X_{3,0}X_{4,0}+X_{3,0}^3)\left(\frac{X_{2,1}-W_{1,1}X_{3,0}}{X_{4,0}-1-X_{3,0}^2}\right)^2\nonumber \\
&& +2\frac{X_{2,1}-W_{1,1}X_{3,0}}{X_{4,0}-1-X_{3,0}^2}\psi(X,X_{3,0},\ldots, X_{N,0})\Phi_3(X_{3,0},\ldots, X_{N,0}) \nonumber \\
&& +\overline{W}_{1,2}(X_{3,0},\ldots, X_{N,0}). \label{eqn:ex3gW12}
\end{eqnarray}
Equations~(\ref{eqn:ex3gWf0})-(\ref{eqn:ex3gW12}) constitute necessary but not sufficient conditions for the Jacobi identity to be satisfied. We remind that the closure problem started with $N-1$ functions of $N+4$ variables, and some of the constraints issued from the Jacobi identity reduced the form of the closures to $N-2$ functions (namely $\widetilde{W}_{N+1,0}$, $\overline{W}_{1,2}$, $\phi_n$) of $N-2$ variables and one function (namely $\overline{\overline{W}}_{3,1}$) of $N-1$ variables. It remains $N-3$ partial differential equations to be solved in order to fully satisfy the Jacobi identity. These equations are not written here since their expression is very lengthy and not necessarily very enlightening. 
We identified a family of solutions $\psi=0$, i.e., when $\overline{W}_{n,1}=0$ for which the resulting system is Hamiltonian for any functions $\widetilde{W}_{N+1,0}$ and $\overline{W}_{1,2}$.  

A major difficulty in the above closures in Eqs.~(\ref{eqn:ex3gWn1})-(\ref{eqn:ex3gW12}) is that they are singular at $X_{4,0}=1+X_{3,0}^2$. This results in a Hamiltonian system which is singular at $X_{4,0}=1+X_{3,0}^2$ There is no way to select appropriately the closure functions $\psi$, $\phi_n$, $\widetilde{W}_{N+1,0}$ and $\overline{W}_{1,2}$ in order to remove the singularities of $\widetilde{W}_{n,1}$ and $\widetilde{W}_{1,2}$. As a consequence, the case for $N\geq 5$ does not lead to non-singular closures.

\section*{Conclusion}

Reducing the number of variables/degrees of freedom is a necessity to reduce computational cost and to derive reduced models to better understand the qualitative phenomena at play. However there is a significant risk in modifying the structure of the parent model, and potentially introduce fake dissipation in the system. We have investigated here some reductions of the one-dimensional Vlasov equation by considering the reductions in the phase-space moments. Using selected examples, we have shown that the requirement that the closure satisfies the Jacobi identity introduces relevant dynamical variables, like for instance, the central moments and their combinations, and drastically affects the resulting dynamics. This is clearly seen by the significant number of invariants created as a result of imposing the Jacobi identity, which shapes the phase space of the resulting reduced models. With the above examples, we have been able to define models with one macroscopic degree of freedom and one or two microscopic degrees of freedom. In all the cases examined in this article, the inclusion of additional moments in the position as dynamical variables did not lead to an increase of the effective dimension of phase space. In order to increase the number of microscopic degrees of freedom, one needs to include higher order moments in both position and velocity.  

\begin{acknowledgments}
The authors acknowledge useful discussions with P.J. Morrison and the Nonlinear Dynamics team of the CPT. 
\end{acknowledgments}

\end{document}